\documentclass[%
 aip,
 amsmath,amssymb,
 reprint,%
]{revtex4-1}

\usepackage{graphicx}% Include figure files
\usepackage{dcolumn}% Align table columns on decimal point
\usepackage{bm}% bold math
\usepackage[utf8]{inputenc}
\usepackage[T1]{fontenc}
\usepackage{mathptmx}
\usepackage{etoolbox}

\makeatletter
\def\@email#1#2{%
 \endgroup
 \patchcmd{\titleblock@produce}
  {\frontmatter@RRAPformat}
  {\frontmatter@RRAPformat{\produce@RRAP{*#1\href{mailto:#2}{#2}}}\frontmatter@RRAPformat}
  {}{}
}%
\makeatother
\begin{document}

\preprint{AIP/123-QED}

\title{{\it Ab-initio} study of ultrafast spin dynamics in Gd$_x$(FeCo)$_{1-x}$ alloys}
\author{J.~K. Dewhurst}
\affiliation{Max-Planck-Institut fur Mikrostrukturphysik Weinberg 2, D-06120 Halle, Germany}
\email{dewhurst@mpi-halle.mpg.de}
\author{S. Shallcross}
\affiliation{Max-Born-Institute for Non-linear Optics and Short Pulse Spectroscopy, Max-Born Strasse 2A, 12489 Berlin, Germany}
\author{I. Radu}
\affiliation{Department of Physics, Freie Universit\"at Berlin, Arnimallee 14, 14195 Berlin, Germany}
\author{P. Elliott}
\affiliation{Max-Born-Institute for Non-linear Optics and Short Pulse Spectroscopy, Max-Born Strasse 2A, 12489 Berlin, Germany}
\author{C. v. Korff Schmising}
\affiliation{Max-Born-Institute for Non-linear Optics and Short Pulse Spectroscopy, Max-Born Strasse 2A, 12489 Berlin, Germany}
\author{S. Sharma}
\affiliation{Max-Born-Institute for Non-linear Optics and Short Pulse Spectroscopy, Max-Born Strasse 2A, 12489 Berlin, Germany}

\date{\today}

\begin{abstract}
Using an ultrashort laser pulse we explore {\it ab-initio} the spin dynamics of Gd$_x$(FeCo)$_{1-x}$ at femtosecond time scales. Optical excitations are found to drive charge from Fe majority $d$-states to the unoccupied Gd $f$-minority states, with $f$-electron character excited occupation lagging behind that of the $d$-electron character, leading to substantial demagnetisation of both species while leaving the global moment almost unchanged. For $x > 0.33$ this results in the creation of an ultrafast ferromagnetic (FM) transient by the end of the laser pulse, with the Gd demagnetization rate slower than that of Fe. For all concentrations the Gd moments begin to rotate from their ground state orientations developing in-plane moments of between 0.2-0.5~$\mu_B$. Thus, the ultrafast spin dynamics of the material captures three important ingredients of the all optical switching that occurs at much later (picosecond) times: (i) the development of a FM transient, (ii) the different rates of demagnetisation of Fe and Gd and, (iii), the breaking of the colinear symmetry of the ground state. Furthermore, several predictions are made about the behaviour of Fe-Gd alloys that can be experimentally tested and can lead to a spin-filtering device. 
\end{abstract}

\maketitle
\begin{quotation}
The field of laser driven spin dynamics on ultrafast time scales (from attoseconds to picoseconds) holds the possibility of profound technological innovation\cite{kirilyuk17,kimel19,radu-book}. Central to this field in both scientific interest and technological importance is a simple question: how can magnetic order be controlled by light? Since the pioneering demonstration of ultrafast demagnetisation in Ni by light more than 20 years ago \cite{Beaurepaire1996}, two purely optical approaches to manipulate long range magnetic order have emerged. At femtosecond time scales the OISTR effect\cite{dewhurst2018} predicts that, in multi-sub-lattice magnetic materials, a transient ferromagnetic state can be created from an uncompensated anti-ferromagnetic ground state. This is a generic early time effect with the long time fate of the transient still uncertain. On the other hand at comparatively much longer time scales, of the order of picoseconds, the direction of spin (i.e. $+z$ or $-z$) in a broad class of materials can be flipped by laser light, the phenomena of all optical switching (AOS)\cite{stanciu_all-optical_2007,vahaplar_ultrafast_2009,radu_transient_2011}. In contrast to the OISTR effect, the switched state achieved in AOS is permanent and not transient. This represents a broad umbrella of different physical processes, including both helicity-dependent\cite{lambert_all-optical_2014,el_hadri_two_2016-1,arora_spatially_2017} AOS (which occurs on slower time scales and only under multiple pulses exposure) and helicity-independent switching\cite{vahaplar_ultrafast_2009,radu_transient_2011,medapalli_role_2013}, which is a single-shot excitation that can be switched back and forth multiple times.
\end{quotation}

%\section{Introduction}
The few femtosecond or sub-femtosecond time scales of the OISTR phenomena ensures that this effect occurs well before either exchange interactions or temperature emerge from the underlying non-equilibrium electronic system as well defined variables. This is exemplified by the theoretically predicted light induced change in magnetic structure from anti-ferromagnetic to ferromagnetic in Co/Mn multilayers\cite{dewhurst2018}, in which the magnetic moments in one layer reverse direction without rotating from the $z$-axis, a phenomena that obviously cannot be described within a Heisenberg model picture of the dynamics of magnetic order.

In Gd based alloys\cite{stanciu_all-optical_2007,chimata_all-thermal_2015} and multilayers\cite{beens_comparing_2019} AOS can be driven by a single intense laser pulse (the polarisation state of which is unimportant) with, however, the switching of Gd moments occurring much after the application of the pulse, of the order of a few picosecond later. In the case of Gd$_x$(CoFe)$_{1-x}$ alloys, an uncompensated anti-ferromagnet in the ground state, the laser pulse results first in reorientation of the Fe moments to create a ferromagnetic transient\cite{radu_transient_2011}, with the Gd spin direction subsequently switching to return the system to the anti-ferromagnetic structure of the ground state, but with all moments now aligned in the opposite direction. Central to this phenomena are believed to be (i) the distinct time scales on which Fe and Gd demagnetise and (ii) the anti-ferromagnetic coupling between Gd and Fe. It should be emphasised that, in contrast to the OISTR effect, the longer time scales of AOS implies that exchange interactions and local moments provide a valid description of the spin dynamics and thus moment rotation may play role in the long time the spin dynamics. In fact, at the experimental stoichiometry simple number of available states argument show that the Gd moment cannot flip by spin-flips or optical excitation induced transitions alone.

These two examples of the manipulation of magnetic long range order by light evidently inhabit very different temporal landscapes: the OISTR effect represents an example of profound spin-charge coupling occurring at sub-exchange times, while AOS in Gd$_x$(CoFe)$_{1-x}$ occurs at picosecond time scales at which not only can the spin dynamics be characterised in terms of exchange interactions but also temperatures reliably assigned to the spin, lattice, and charge sub-systems. The question we wish to address in this work is to what extent these two different pictures of spin dynamics are connected: what role if any does the early time (purely electronic) spin dynamics have in the phenomena of AOS in Gd$_x$Fe$_{1-x}$? In particular, could the ferromagnetic transient be partly an early time OISTR phenomena?

By doing fully \emph{ab-initio} calculations we show (i) that the OISTR effect can create a ferromagnetic transient already at $\sim 10$ femtoseconds, (ii) that increasing the concentration of Gd results in larger change in moment of Fe, but concomitantly a reduced demagnetisation of Gd, (iii) that the excitation of the Gd $f$-electrons lags behind that of the $d$-electrons mostly likely indicating an indirect Fe-$d$ to Gd-$f$ transitions via Gd-$d$ states, (iv) the symmetry (i.e. the collinear magnetic alignment)  is broken at very early times, with the Gd moments already at 20~fs exhibiting out-of-plane components, and (v) the individual Gd moments behave incoherently i.e. they all evolve very different in-plane components. Strikingly therefore, two of the key ingredients of the later time AOS -- the existence of a ferromagnetic transient and the breaking of collinearity -- already have their footprint at very early times, suggesting the possibility of new routes to control AOS in this material.

%\section{Methodology}

In the present work fully \emph{ab-initio} calculations are performed comprising of two steps: (a) the ground-state is first calculated using density functional theory (DFT) using for the exchange-correlation (xc) functional the local spin density approximation together with Hubbard $U$ (LSDA+$U$) and subsequently (b) the time dependent extension of density functional theory (TD-DFT)\cite{RG1984,my-book} is employed as a fully first principles approach to study the dynamics of charge and spins. For this we use adiabatic LSDA+$U$ as the xc functional. The use of the LSDA+$U$ functional for both ground state and dynamics is required in order to correctly treat the highly localized Gd $f$-states; we find a value of $U=6.7eV$ applied on the Gd $f$-states leads to the correct positioning in energy of the occupied and unoccupied $f$-states. This value of $U$ was kept constant during the time propagation.  

This first principles approach for the ultrafast dynamics of spin and charge has been shown to accurately describe magnetic systems on femtosecond time scales\cite{dewhurst2018,siegrist2019,steil2020,hofherr2020,clemens2020,chen2019}. A time step of 2.4 attoseconds was used for time propagation (see Ref.~\cite{dewhurst2016} for full detail of the method). All calculations were performed within the framework of the all-electron full-potential linearized augmented plane wave method\cite{singh}, as implemented in the Elk code\cite{elk,dewhurst2016}.

We have used an ultra-short laser pulse of full width half maximum (FWHM) 3~fs. While the FWHM of this pulse resides within the range of experimentally feasible pulses, the incident fluence (88~mJ/cm$^2$) and peak power (4$\times10^{13}$~W/cm$^2$) would appear to be well beyond the threshold for maintaining structural integrity of the sample. However, it is important to note that in our calculations $\sim80$\% of the absorbed energy is used to excite to highly excited states. In experiments  these states have very short lifetimes, decaying within a few femtoseconds and thus releasing their energy for excitation of the lower energy states which are longer lived. The theoretical and experimental fluences thus cannot be directly compared, as in theoretical case the infinite life time of excited states precludes this energy transfer from high to low energy states that occurs in experiment. Therefore a theoretical fluence $\sim80$\% above the typical experimental values corresponds, in terms of the energy absorbed by the important low energy states, to a reasonable experimental fluence.

We have used the C15 cubic Laves (spacegroup 227) unit cell with lattice parameter of 7.32~au. This unit cell contains 6 atoms and thus allows for the study of concentration dependence via variation of the number of Gd and Fe atoms (e.g. 33\% Gd and 67\% Fe unit cell consists of 4 Fe and 2 Gd atoms). A 12$\times$12$\times$12 {\bf k}-point set was used and conduction states up to 50~eV above the Fermi energy were included.

%\section{Results}

\begin{figure}[t!]
\includegraphics[width=0.9\columnwidth, clip]{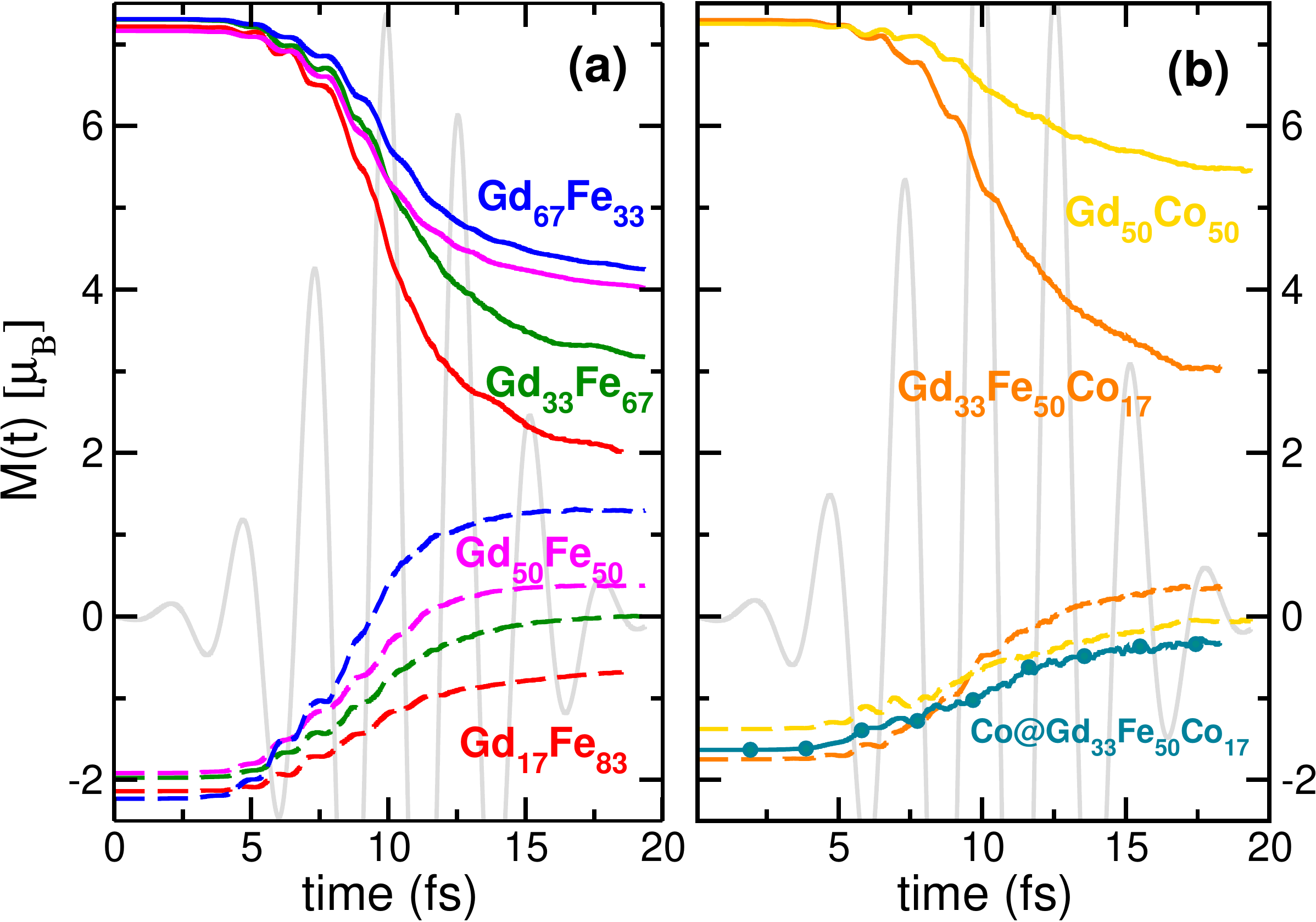}
\caption{Moment on Gd (solid lines) and Fe (dashed lines) in (a) Gd$_x$Fe$_{1-x}$ alloys, (b) Gd$_{33}$Co$_{17}$Fe$_{50}$ and Gd$_{50}$Co$_{50}$ alloys. The laser pulse is also shown in grey and has pulse parameters full width half maximum 3~fs, frequency 1.55~eV, peak power 4x10$^{13}$~W/cm$^2$, and fluence 85~mJ/cm$^2$. Note that only the $z$ component of the moments is shown here; the laser pulse also induces an out-of-plane component, see Figs.~3 and 4.}\label{conc}
\end{figure}

In Fig.~1a we present the time dependent moments of Gd and Fe in Gd$_x$Fe$_{1-x}$ for four different values of $x$ under the influence of a pump laser pulse. Note that only the $z$-component of the magnetisation is shown, and that in the ground state all moments are aligned along the $z$-axis (with the exception of the highest Gd concentration for which the ground state exhibits a small canting of the Gd moments). As can be seen, for all concentrations the laser pulse results in significant demagnetisation of both Gd and Fe, with the maximum decrease in Gd moment obtained at the highest Fe concentration and, vice versa, the maximum change in Fe moment occurring for the highest Gd concentration. At the highest Gd concentration ($x=0.67$) the Fe demagnetizes and re-magnetises in the opposite direction with a moment of $1.6\mu_B$, to create a transient ferromagnetic state. On the other hand for Gd concentrations close to those at which AOS switching occurs in experiment ($x=0.33$) the Fe lattice is already fully demagnetized at the end of the pulse.

It should be stressed that due to their opposite alignment the change in total moment is comparatively small, with in all cases the change in global moment is less than 15\%. This spin dynamics of profound change in local moments alongside a comparatively small change in the global moment rules out spin-orbit coupling as driving the demagnetisation and is characteristic of the OISTR effect that dominates early time spin dynamics in multi-component magnets. This reflects the time scales of the key underlying physical processes: the purely optical excitations that drive the OISTR effect occur on a much faster time scale than spin-orbit induced transitions between majority and minority channels, which are yet faster than the characteristic time coupling of spin angular momentum to the lattice degrees of freedom. The demagnetisation of Fe and Gd, for the ultrashort laser pulse we consider, would thus appear to be dominated by the OISTR effect.

\begin{figure}[t!]
\includegraphics[width=0.9\columnwidth, clip]{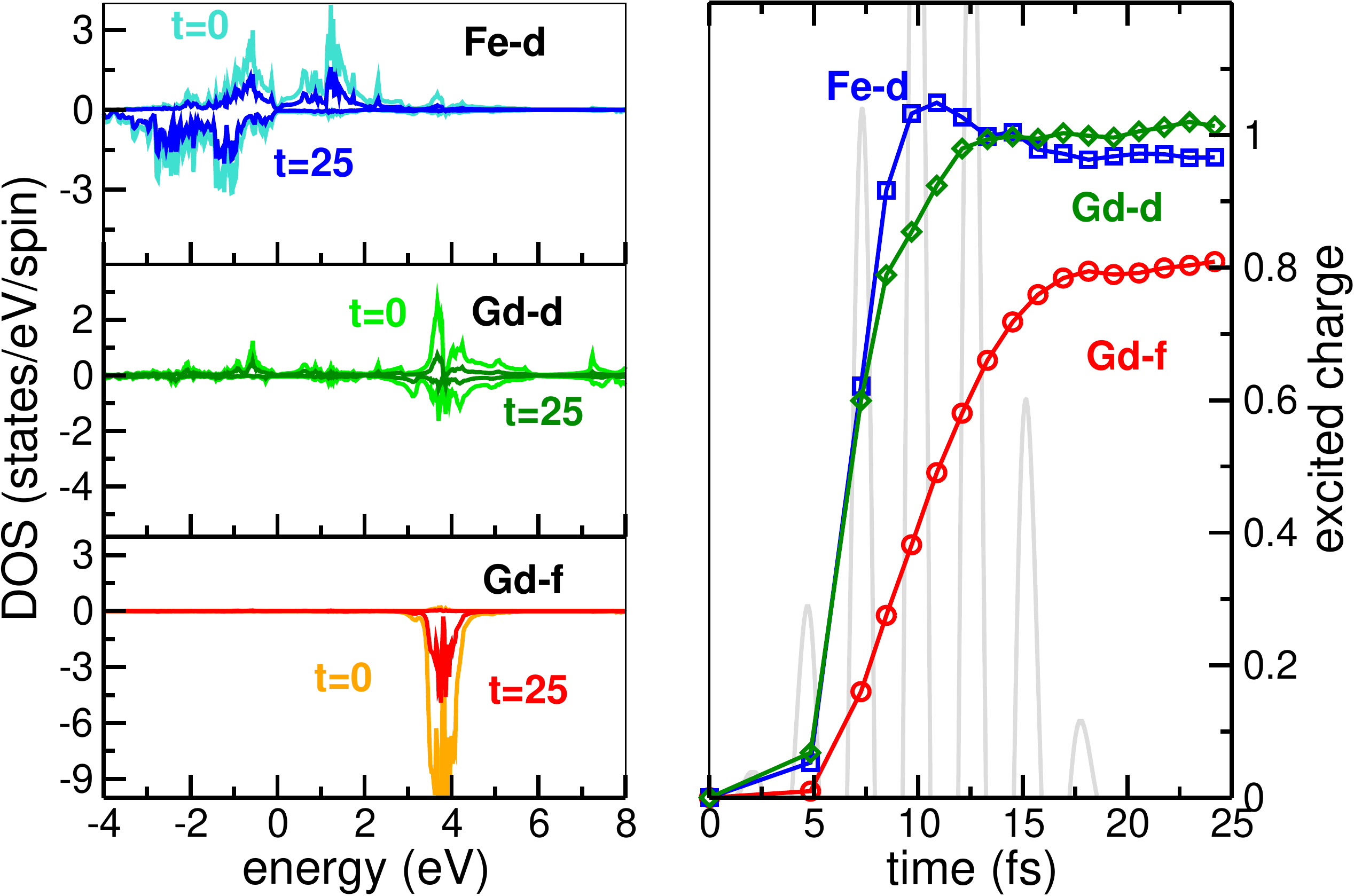}
\caption{(a) Species and angular momentum projected, ground-state DOS and occupied DOS at t=25~fs (b) excited charge projected on the Gd and Fe states. For both cases the system shown is a Gd$_{50}$Fe$_{50}$ alloy. In panel (b) the laser pulse is also shown in grey and has full width half maximum 3~fs, frequency 1.55~eV, peak power 4x10$^{13}$~W/cm$^2$, and fluence 85~mJ/cm$^2$}\label{dos}
\end{figure}

To confirm this in Fig.~2a we show the density of states for  Gd$_{50}$Fe$_{50}$ both in the ground state (labelled by $t=0$) and at $t=25$~fs after the action of the pulse. The ground state DOS shows, as is usual, both occupied and unoccupied states with all states above $E_F=0$~eV (indicated by the vertical line) being unoccupied, while the transient DOS shows only occupied states. As can be seen there is substantial depletion of both spin channels in Fe, however the minority channel is largely inactive in the dynamics of the magnetic moment: optical excitations merely promote minority spin charge from below to above $E_F$ without changing the moment. The behaviour of the Fe majority channel, however, is dramatically different: optical excitations create transitions from majority Fe $d$-states to minority Gd $f$-states. This transition from majority to minority within the same spin channel is a consequence of the anti-ferromagnetic alignment of the Fe and Gd moments, and the net result is to reduce the moment on both Gd and Fe, exactly as is seen in the dynamics of the moment presented in Fig.~1a. Note that while there is some excitation from Fe into Gd $d$-states (see Fig.~2b) -- as transitions from Fe to Gd occur in both majority and minority channels this is less important for the local Gd moment, but does contribute to the reduction of the local Fe moment. Most interestingly, as can be seen in Fig.~2b the Gd $f$-state charge excitation discernibly lags behind that of Fe $d$ and Gd $d$ charge excitation, with at the pulse peak the Fe and Gd $d$-charge having reached close to their final values, but the Gd $f$ character charge less that $\sim50$\% of its final value. This may indicate that the charge in Gd-$f$ states is excited via Gd-$d$ states due to hybridisation; this process of indirect excitation is likely driven by the comparatively small optical matrix element between Fe-$d$ and Gd-$f$ states.

The OISTR effect depends crucially on the number of states available to be excited and the number of empty states available to be excited into. Increasing Gd concentration creates a greater number of unoccupied minority states, thus leading to a greater change in the Fe moment upon laser excitation but concomitantly less reduction of Gd. Correspondingly, at high Fe concentrations the number of occupied majority Fe $d$-states increases, leading both to increased Gd demagnetisation on laser excitation but less loss of the average Fe moment. These "state counting" arguments (i.e. number of available states) evidently can also be extended to changing the alloy species. Substitution of Co for Fe in this way, due to the reduced exchange splitting in Co, is expected to lead to less moment loss on laser excitation. This is shown in Fig.~1b where in comparison with Gd$_{50}$Fe$_{50}$ it is seen that, as expected, in Gd$_{50}$Co$_{50}$ both species show less change in their magnetic moment. The dependence of the early time spin dynamics on Fe concentration and Co substitution are theoretical prediction that can directly checked in experiments.  

\begin{figure}[t!]
\includegraphics[width=0.9\columnwidth, clip]{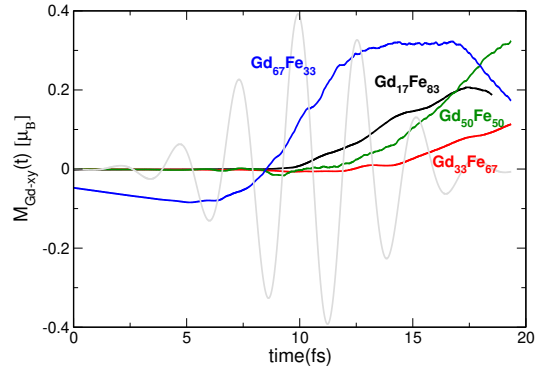}
\caption{Average moment for Gd atoms in $x$-$y$ plane for 4 different concentrations of Gd$_x$Fe$_{1-x}$ ordered alloy. For the largest Gd concentration the ground state already exhibits non-collinearity of the Gd moments due to the long range RKKY character of the Gd-Gd coupling. For all concentrations the Fe local moments are aligned along the $z$ axis, both in the ground state and after the laser pulse.}\label{perpmom}
\end{figure}

We now consider the question of the time evolution of the in-plane components of the magnetic moments. Shown in Fig.~3 is moment in the xy plane averaged over all the Gd atoms in the cell, i.e $\overline{m_{xy}} = \frac{1}{N}\sum_{i=1}^N \sqrt{m_{i,x}^2 + m_{i,y}^2}$ with $N$ the number of Gd atoms and $m_{i,x}$, $m_{i,y}$ the $x$ and $y$ components of the $i$'th atom. As can be see, except for the case of the highest Gd concentration, Gd$_{67}$Fe$_{33}$, a clear in-plane moment develops starting at approximately the time of peak of the pulse (shown in grey), at 10~fs. For Gd$_{67}$Fe$_{33}$, the Gd moments are already slightly canted from the $z$ axis, with a finite in-plane moment observed at $t=0$. Such a non-collinear character ground state is expected for high Gd concentration as the Gd exchange interactions are both weak in magnitude and frustrated due to their long range RKKY character. In this case the average in-plane moment starts to rotate at the same time of $\sim 10$~fs seen for the other concentrations. In contrast the Fe local moments we find, for all alloy concentrations, to be always aligned along the $z$-axis, both in the ground state and during the spin and charge dynamics.

\begin{figure}[t!]
\includegraphics[width=0.95\columnwidth, clip]{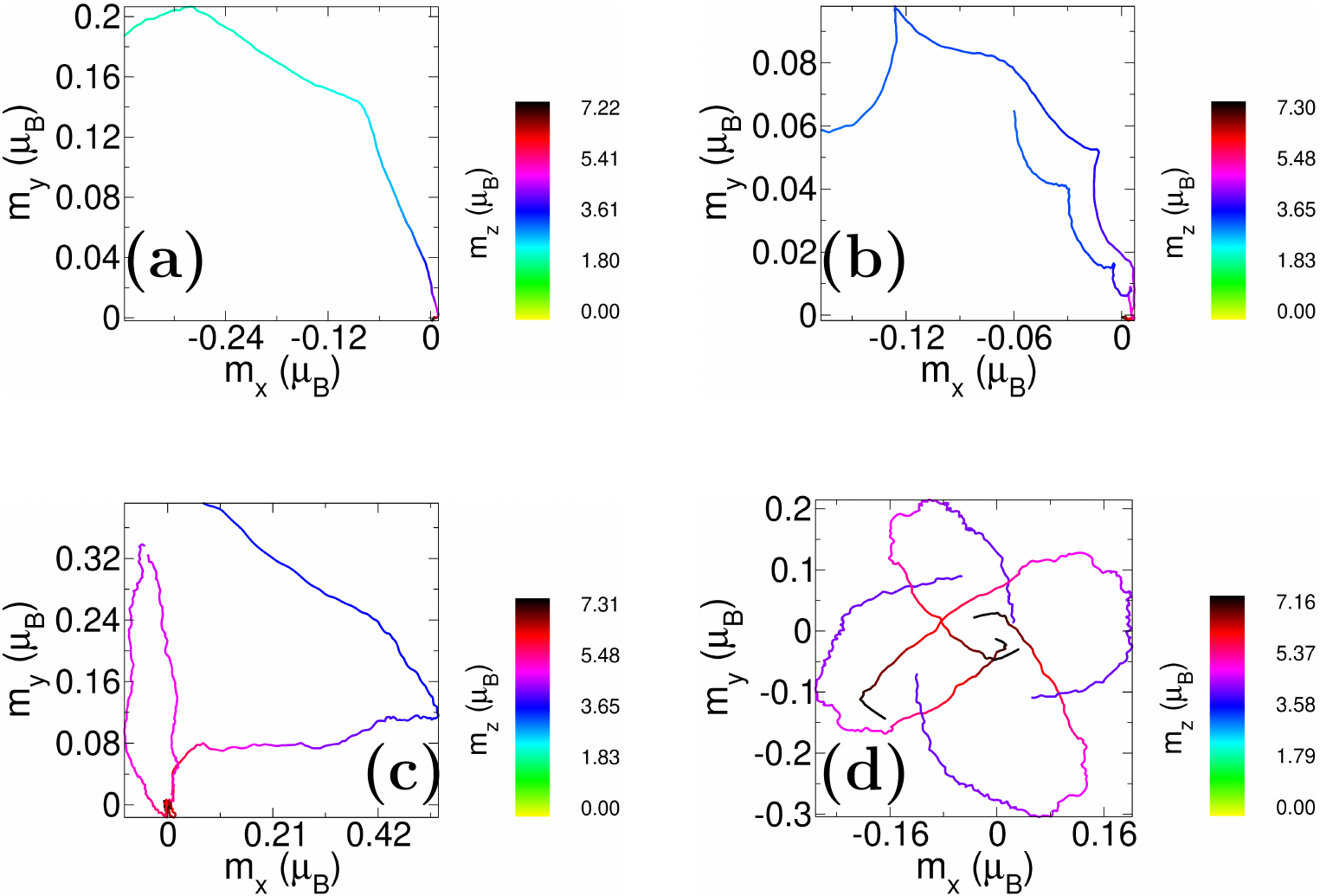}
\caption{Moment on various Gd atoms in xy plane for 4 different concentrations of Gd$_x$Fe$_{1-x}$ ordered alloy; (a) Gd$_{17}$Fe$_{83}$, (b) Gd$_{33}$Fe$_{67}$, (c) Gd$_{50}$Fe$_{50}$, and (d) Gd$_{67}$Fe$_{33}$. Note that the Gd local moments all move in different directions under laser excitation, rendering it difficult to probe them via angle dependent spectroscopy. The Fe local moments, in contrast, are aligned with the $z$-axis both in the ground state and under laser excitation.}\label{xymom}
\end{figure}

An important question is whether the out-of-plane moment develops in a coherent manner i.e. if the canting from the $z$-axis is the same for all Gd atoms. If this were the case, then an angle dependent XMCD signal would be able to measure this.
To better visualise the development of the in-plane components in Fig.~4 we show the evolution of the in-plane components for each Gd atom, with the colour indicating the size of the moment in the $z$ direction. Shown in panel (a-d) are, respectively, Gd$_{17}$Fe$_{83}$, Gd$_{33}$Fe$_{67}$, Gd$_{50}$Fe$_{50}$, and Gd$_{67}$Fe$_{33}$. These have 1, 2, 3, and 4, Gd atoms in the 6 atom unit cell that we employ. Evidently, these moments behaves quite incoherently, with each Gd atom developing its own in-plane components. This early time rotation of the Gd moments from the $z$-axis will therefore not be measurable in any element-specific magnetic techniques that average over macroscopic laser-excited areas/volumes, e.g. XUV/soft X-rays, XMCD, and MOKE.

%\section{Conclusions and outlook}

To summarize, using state-of-the-art {\it ab-initio} time dependent density function theory we have addressed the early time spin dynamics of ordered Gd$_x$(CoFe)$_{1-x}$ alloys driven by a short pulse of 3 femtosecond full width half maximum (FWHM), a duration comparable to the currently shortest available pulses available in experiment. We find that the early time spin dynamics is dominated by OISTR physics: optically induced excitations from Fe-$d$ majority to Gd minority result in a substantial loss of moment from both species, while leaving the global moment almost unchanged. The smaller moment of Fe nevertheless results in a much faster demagnetisation of this species, with for concentrations $x > 0.33$ re-magnetisation in the opposite direction creating a ferromagnetic transient. We also find that the Gd $f$-electron character excited charge lags behind the $d$-electron excitation of Gd and Fe: at the pulse maximum the latter are already close to the final values with, on the other hand, the excited charge of Gd $f$-electron character having reached only $\sim50$\% of its final value. This might indicate an indirect transfer of charge from Fe-$d$ states to Gd-$f$ states via Gd-$d$ states, a process made more likely by the comparatively small optical matrix element for direct Fe-$d$ to Gd-$f$ transitions.

The early time spin dynamics also displays clear symmetry breaking of the collinear ground state, with small but significant rotation of the Gd moments occurring within the trailing edge of the pulse (0.2-0.5$\mu_B$ in-plane components). Thus three key aspects of the later time spin dynamics in all optical switching (AOS) in Gd$_x$Fe$_{1-x}$ alloys are thus found in the very early time regime: (i) the FM transient, (ii) the slower rate of demagnetization of Gd as compared to Fe and, (iii), the breaking of ground state collinearity.

The magnitude of the OISTR effect depends on both on the available empty states as well the states available for optical transitions to excite into this unoccupied band. The moment of the FM transient, as well as the magnitude of Gd demagnetisation, can thus be tuned both by alloy concentration and alloy species e.g. by substituting Fe by Co, and  our calculations predict that the early time moment loss can be controlled by substitution of Fe by Co. Furthermore, the time scale on which this transient is created will be dictated solely by pulse parameters. We thus find that the FM transient in Gd$_x$(CoFe)$_{1-x}$ is in principle highly controllable in the early time regime. The dependence of the early time spin dynamics on Fe concentration and Co substitution are two key \emph{ab-initio} predictions that can be investigated in experiments. Interestingly, while we have only considered ordered alloys the OISTR effect has previously been demonstrated to hold in an interface geometry\cite{substrate18,chen2019}. In this case spin currents across the interface result in changes to moments of atoms adjacent to the interface. Recent experiments demonstrating all optical switching in a multilayer geometry are thus consistent with the early time picture we study here.

The transient FM state of Fe-Gd also implies that this material is an excellent choice for the a so-called OGMR effect (OISTR induced Giant Magnetic Resistance). The material being initially ferri-magnetic with oppositely aligned Gd and Fe moments implies that if a spin polarised current was to pass through such a material, both spin channels would feel a large resistance. However, during the laser induced transient FM state, one of the spin channels (parallel to the direction in which the FM spins are oriented) feels much less resistance. Hence such a setup can act as a transient spin-filter. This is again a prediction which can be experimentally directly tested and could lead to a transient spin-filtering device.

For the longer pulses typically employed in the study of AOS the OISTR physics described here is likely to be of importance, as such physics plays an important role at such time scales in other materials, augmented by lattice and SOC effects that become more significant at 100s of femtoseconds. How this highly non-equilibrium early time regime evolves into a regime in which the  effective exchange interactions and temperature emerge as good variables remains perhaps the most significant unanswered question for AOS. Indeed, all present theories of AOS address this later time domain and cannot describe early time electronic effects such as the OISTR physics demonstrated here. Future work to bridge these two time regimes and their distinct theoretical methodologies, perhaps with {\it ab-initio} calculations providing "initial conditions" for models of later time dynamics, might be expected to lift many of the uncertainties surrounding the origins of this remarkable effect.

\section*{Acknowledgements}

Sharma, CvKS and IR would like to thank DFG for funding through TRR227 (project A04, A02 and A05). Shallcross would like to thank DFG for funding through SH498/4-1 while PE thanks DFG for funding through DFG project 2059421. IR also acknowledges funding from European Research Council through TERAMAG (Grant No. 681917). The authors acknowledge the North-German Supercomputing Alliance (HLRN) for providing HPC resources that have contributed to the research results reported in this paper.

\section*{Data Availability Statement}
The data that support the findings of this study are available from the corresponding author upon reasonable request.

%\bibliography{Gd}
%merlin.mbs aipnum4-1.bst 2010-07-25 4.21a (PWD, AO, DPC) hacked
%Control: key (0)
%Control: author (8) initials jnrlst
%Control: editor formatted (1) identically to author
%Control: production of article title (0) allowed
%Control: page (1) range
%Control: year (1) truncated
%Control: production of eprint (0) enabled
%

\end{document}